# Intense THz s-SNOM for nonlinearity engineering in nanoscale

Pengfei Qi[1,2,3,#], Zeliang zhang[1,2,3,#], Wenqi Qian[1,2,3], Zijie Dai[1,6], Xingyou Li[1,2,3], Lu Sun[1,2,3], See Leang Chin[2,4], Pierre Agostini[2,5], Weiwei Liu[1,2,3*]

[1] *Institute of Modern Optics, Eye Institute, Nankai University, Tianjin 300350, China*
[2] *Pierre Agostini International Joint Research Center for Ultrafast Optics and Applications, Nankai University, Tianjin 300350, China*
[3] *Tianjin Key Laboratory of Micro-scale Optical Information Science and Technology, Tianjin 300350, Chinas*
[4] *Department of Physics and Center for Optics, Photonics and Lasers (COPL), Laval University, Quebec G1V 0A6, Canada.*
[5] *Department of Physics, the Ohio State University, Columbus, Ohio 43210, USA*
[6] *Institute of Micro-nano Optoelectronics and Terahertz Technology, Jiangsu University, Zhenjiang, Jiangsu 212013, China*
* *liuweiwei@nankai.edu.cn*
[#] *Pengfei Qi and Zeliang Zhang contributed equally to this work.*

**Abstract:** Terahertz (THz) nonlinear optics offer powerful tools to investigate and manipulate electronic dynamics in condensed matter. Confining high-peak-power THz pulses within near field can effectively generates extremely localized electromagnetic fields in spatio-temporal, enabling to precisely explore and control carrier transient dynamics from THz nonlinearity perspective. However, the combination of the high peak power THz pulses and the near-field optic techniques remains challenging due to the incompatibility between low repetition THz pulses and typical near-field demodulation schemes. Here, we construct high peak power THz scattering scanning near-field microscopy (THz s-SNOM) by combining THz pulses emitted from two-color femtosecond laser filaments with a tapping mode atomic force microscopy (AFM) and explore efficient THz third harmonics generation (THG) from the $Cd_3As_2$ film in nanoscale. The power-law dependence of the THz harmonics and theoretical calculation reveals a convincing third harmonic generation that is attributed to the nonequilibrium intraband dynamics driven by the strong THz pulses. Especially, the nanoscopic near-field THz third harmonic imaging with resolution of 200 nm ($\lambda$/3000) of 3D Dirac semimetal are demonstrated. The high peak power THz s-SNOM can provide a great platform for exploring and manipulating the nonlinear physics, carrier dynamics and quantum coherent phenomena driven by the localized THz field with nanoscale resolution, thereby guiding the development of the integrated high-performance nonlinear photonic devices.

## INTRODUCTION

Nonlinear optics (NLO) has grown substantially in past decades, leading to tremendous progress in fundamental research and revolutionized applications[1–3]. Recently, due to the great demand exists in modern ultrahigh-speed electronics for efficient active functional materials supporting very high terahertz ($10^{12}$ Hz, THz) frequencies[4–6], THz NLO has emerged and provided powerful probes and tools to examine and control the electronic dynamics of crystalline solids, such as dynamical Bloch oscillations, quantum interference and intra-band current[7–9]. Currently, THz nonlinearity was generally excited in macroscopic spatial scale (~mm) by the strong THz pulses (~100kV/cm). Although the latest reports reveal the ultrafast dynamics of carriers during the THG process, capturing the THz nonlinearity modes in nanoscale is still a very demanding task[10–13], which is critical for exploring the quantum confinement effect on nonlinear optical physics and developing the integrated high-performance nonlinear photonic devices.

Noteworthily, the THz s-SNOM, coupling THz wave with the nanotip, has been adopted to probe the mobile carrier oscillations (plasmons) and lattice vibrations (phonons) on the nanoscale[14–16]. The nanoscale sharp metal AFM tip can confine the electromagnetic field of the incident THz wave within nanoscale in spatial, far beyond the diffraction limit, and greatly enhances the interaction between THz wave and matter[17–19]. Generally, the THz s-SNOM is always built on a continuous or high-repetition THz source to satisfy the near-field demodulation scheme. The locally enhanced continuous or high-repetition THz wave is still difficult to surpass the peak intensity of strong THz pulses to induce nonlinear processes[20–23]. Therefore, the nonlinear phenomena in nanoscale excited by THz s-SNOM has not been reported yet. Naturally, coupling ultrafast and strong THz pulses to nanotip apex provide a preferred route for investigating and modulating the nonlinear optical physics at nanoscale[24–26]. However, the near-field demodulation scheme for SNOM requires the repetition rate (~10 MHz) of the incident THz wave to be much higher than the mechanical oscillation frequency (~10 kHz) of the nanotip. Therefore, operating the THz s-SNOM with a high-peak-power and low-repetition THz source to excite near-field nonlinear generation remains a scientific challenge[27–29].

Here we report an intense THz s-SNOM for nonlinearity engineering in nanoscale by combining high peak power THz pulse emitted from two-color femtosecond laser filaments with common THz s-SNOM. Firstly, the efficient THz THG in three-dimensional Dirac semimetal (3D DSM) $Cd_3As_2$ with nanoscale localized field enhancement was demonstrated by the intense THz s-SNOM. The nonequilibrium intraband dynamics driven by strong THz pulses near the 3D Dirac cone of $Cd_3As_2$ was conducted to describe the near-field THz nonlinearity. The power-law dependence of the THz harmonics and theoretical calculation reveals a convincing third harmonic generation that attributed to the nonequilibrium intraband dynamics driven by the strong THz pulses. Secondly, the near-field third harmonic imaging with resolution of 200 nm ($\lambda$/3000), that is nanoscopic near field THz THG of 3D Dirac semimetal are demonstrated. Especially, the broadband THz pulse emitted from filaments support to produce tunable narrow-band THz wave via filter, which is conducive to selectively and resonantly excite the fine electronic band structure for precisely characterizing and controlling the non-equilibrium dynamics in novel materials. These findings can provide insights for nonlinear currents of Dirac fermions driven by the localized THz field and deepen our understanding of the nonlinear interaction between Dirac semimetal and THz waves in near field, illustrating the high peak power THz s-SNOM can provide a great platform for exploring and manipulating the nonlinear physics and carrier dynamics driven by the localized THz field with nanoscale resolution.

## RESULT AND DISCUSSION

**THz third harmonics generation from $Cd_3As_2$ thin film in near field.**

Our custom-made THz s-SNOM is based on a tapping-mode atomic force microscope (AFM) excited by high peak power THz pulses generated by the two-color laser filament (Figure 1a). The AFM nanotip acts as an antenna that receives far-field THz pulses, amplifies the near-field interaction through an enhancement of the THz field by tip resonances, and finally transmits and scatters the THz near-field to the far field. Note that the typical repetition rates of 1 kHz for such THz pulses are significantly lower than the tip oscillation frequency, here we introduce a synchronous sampling method (SSD) in THz s-SNOM to capture the weak scattering THz signals. Specifically, controlling the nanotip approaches to the sample surface when the THz pulse arrives (Figure 1b), thereby the near-field scattering signal can be synchronously and effectively generated. That is, the repetition frequency of near-field signal matches well with the vibration frequency of nanotip, creating a near-field scattering THz pulse modulated by the vibration frequency of nanotip, which can be reliably decoupled from the background noise.

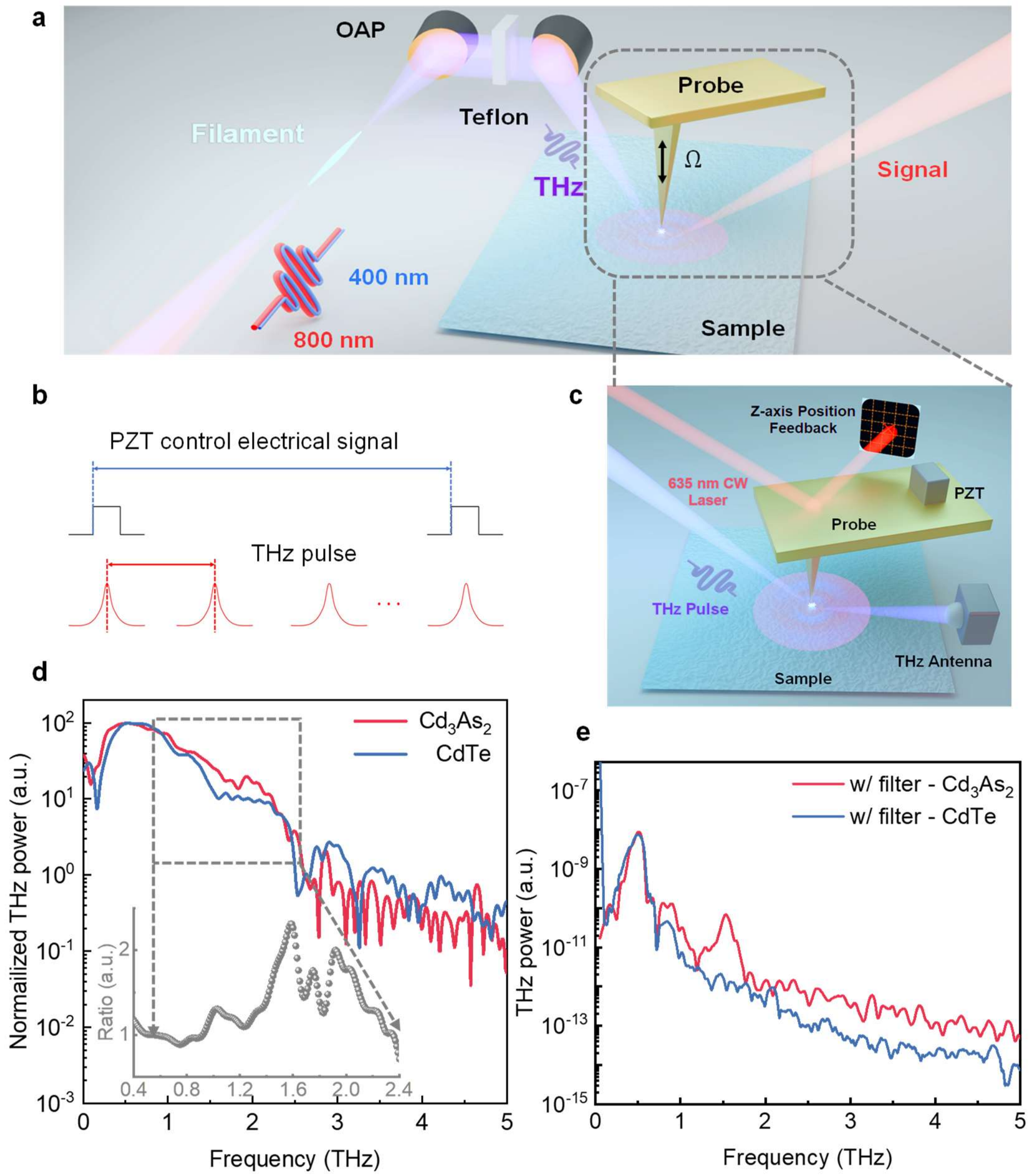

**Figure 1. THz third harmonics generation from $Cd_3As_2$ thin film in near field.** a. Diagram of intense THz s-SNOM combining with high peak power THz pulse emitted from two-color femtosecond laser filaments. b. Illustration of the THz pulse, piezoelectric ceramic vibration, and photoconductive antenna maintain sequential synchronization. c. THz pulse is coupled to the near-field system composed of the tip apex and the sample. The nanotip cantilever is controlled by a piezoelectric ceramic. A 635 nm laser is used for real-time detection of the probe's position in the Z-direction. d. Near-field spectra were measured without a 0.5 THz filter. Inserted figure: difference between signal from $Cd_3As_2$ and CdTe. e. Near-field scattering spectra with a 0.5 THz filter.

As illustrated in Figure 1c, the high peak power THz pulses with broadband spectra and linearly polarization illuminate the gap between Al-coated nanotip with radius of 10 nm and the 3D DSM $Cd_3As_2$ film to excite and detect the near-field THz nonlinearity, where the nanotip cantilever is controlled by a piezoelectric ceramic. Note that the 3D DSM, also termed “3D graphene,” has attracted great attention due to the exotic condensed-matter physics phenomena and great potential in optoelectronics[30–32]. Compared with monolayer graphene, 3D DSM exhibits linear band dispersion while having a macroscopic thickness, which can substantially enhance interaction volume compared with graphene, thereby provide an additional degree-of-freedom to improve THz nonlinearity efficiency[33,34]. As a typical 3D DSMs, the $Cd_3As_2$ with stable structural feature, massless charge carriers and large linear energy momentum space, demonstrates an extreme nonlinearity in the THz regime[35–37]. Currently, due to the challenging in the near-field manipulation of strong THz waves, 3D DSM based THz nonlinearity was also generally excited in macroscopic spatial scale. Here the thin $Cd_3As_2$ films, with a thickness of 100 nm, are grown epitaxially on a sapphire substrate with a CdTe buffer layer. The energy band structure of $Cd_3As_2$ features two Dirac nodes near the $\Gamma$ point and exhibits quasi-linear band dispersion up to ~1 eV. For the low carrier concentration and effective mass, the Fermi energy is estimated to be at a minimum ~50 meV above the Dirac point.

Then we performed systematic measurements utilizing the above THz s-SNOM to evaluate the nonlinear response of $Cd_3As_2$. The broadband THz pulses with/without narrowband filter were adopted to characterize the near-field nonlinear optical response of the $Cd_3As_2$. The peak electric field intensity of the THz radiation generated by the two-color laser filamentation is approximately 300 kV/cm (extended Figure S1). After propagating through the filter, the peak electric field intensity of the narrow-band THz wave with a central frequency of 0.5 THz is approximately 60 kV/cm (See the methods for more details). For the distance between the nanotip apex and the sample surface of 50 nm, the localized peak THz electric field strength at the nanotip apex is ~350 kV/cm, enhanced by localized field effect (See Equations. 6-7 and extended Figure S7 for more details). Figure 1c illustrates near-field THz scattering spectra of the $Cd_3As_2$ film and CdTe substrate, illuminated by the original broadband THz pulses, respectively. Note that the photon energy of fundamental wave is much lower than the band gap of CdTe (approximately 1.44 eV), therefore the electron dynamics and nonlinearity in CdTe cannot be excited by the incident THz field. Clearly, the near-field scattering spectra at 1~2 THz of $Cd_3As_2$ was significantly enhanced compared with the CdTe substrate, revealing the generation of new spectral components (Inset of Figure 1c, ratio between the signals corresponding to $Cd_3As_2$ film and CdTe substrate), due to the induced THz near-field nonlinearity by high peak power THz pulses. However, the related nonlinear mechanism is very difficult to be distinguished because of the broadband fundamental THz waves.

To exhibit the nonlinear effect more clearly, the filter was subsequently inserted to obtain the narrowband fundamental THz waves (the peak electric field intensity of the narrow-band pump THz field is 60 kV/cm). Selecting the appropriate driving field frequency is crucial for enhancing the nonlinear conversion efficiency. For the Dirac semimetal, nonlinear electron dynamics are governed by the acceleration theory, which demonstrates a positive correlation between nonlinear conversion efficiency and conductivity. It was demonstrated that the conductivity of $Cd_3As_2$ is both tunable and sensitive to the pump fluence for THz waves ranging from 0.4 to 1.5 THz. Additionally, the real component of the conductivity reaches maximum at 0.5 THz, while the imaginary component is exactly minimum.[38] Consequently, here the filter with central frequency at 0.5 THz was chosen to optimize the conductivity and improve the nonlinear conversion efficiency. As we expected, the eye-catching THz third harmonics generation centering at 1.5 THz are observed in near-field THz scattering spectra for the $Cd_3As_2$ film excited by the THz pulse with central frequency at 0.5 THz (Figure 1d, red line), while there is no third-harmonic generation can be found in the CdTe film (Figure 1d, blue line). It can be concluded that the localized field of strong THz pulse induced by nanotip can significantly improve the nonlinear harmonic conversion efficiency of 3D Dirac semimetal (DSM) at THz frequency.

Considering the scattering efficiencies of fundamental wave (0.5 THz) and THG (1.5 THz) have no significant differences (see extended Figure S5), the nanotip-$Cd_3As_2$ near-field scattering spectra can be directly used to estimate the third-order THz nonlinear optical coefficients of $Cd_3As_2$. Figure 2 illustrates the extracted light filed in temporal of fundamental wave (0.5 THz) and THG (1.5THz). Note that the pump field is localized in a nanoscale of < 200 nm and there is approximately 90% field transmission through the $Cd_3As_2$ film with a thickness of only 100 nm. Therefore, the phase mismatch and the pump field depletion can be ignored during THG process. The conversion from the pump to the harmonic field can be defined as $\left|E_{mf}\right| = \gamma L \left|E_f\right|^m$ with $\gamma = k_m\left[\pi f/\left(cn_{mf}\right)\right]\chi_{eff}^{(m)}$ [39]. Here $m$ is the harmonic order; $E_f$ and $E_{mf}$ are peak electric field of the fundamental and $m$ th harmonic field; $k_m$ = 3/4 is a numerical coefficient for third harmonic; $L$ is the thickness of the film; $c$ is the speed of light in vacuum. From the measured THG with $f$ = 0.5 THz → $3f$ = 1.5 THz, the field conversion coefficient of $\gamma L = \left|E_{3f}\right|/\left|E_f\right|^3 = 1.45\times10^{-16} m^2V^{-2}$, and the effective third-order THz nonlinear susceptibility of $\chi^{(3)}_{eff} \approx 1.4\times10^{-9} m^2V^{-2}$ were estimated, which is in accordance with the previous results of structurally analogous Dirac materials[33,39]. The near-field scattering efficiencies for different frequencies were also considered in the calculation to guarantee accuracy.

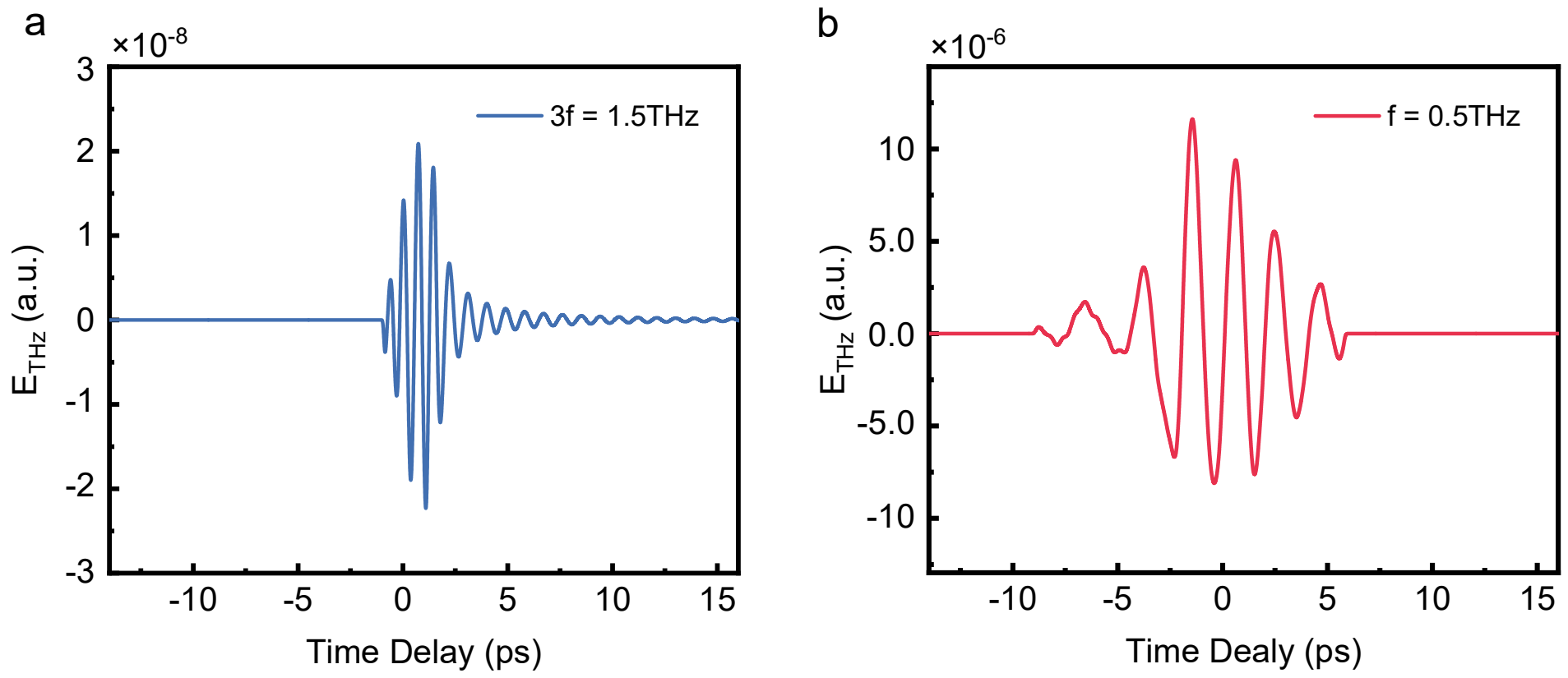

**Figure 2 Near-field scattering signal from nanotip apex – $Cd_3As_2$ in temporal.** a Third harmonic time domain curve. b Fundamental wave time domain curve. The third harmonics were Fourier-filtered from the complete nonlinear signal containing fundamental wave and the third harmonic at once.

## Nonequilibrium intraband dynamics driven by THz pulses

To understand the above near-field THz nonlinearity generation, nonequilibrium intraband dynamics of electrons driven by strong THz pulses near the 3D Dirac cone of $Cd_3As_2$ were performed. For interband electronic excitations at room temperature, one-photon transitions within the THz frequency band are forbidden by Pauli-blocking due to the thermally excited electron doping. Consequently, here the nonequilibrium intraband dynamics of Dirac electrons were emphatically calculated. The initial state of thermodynamic equilibrium at room temperature for electrons obey the Fermi-Dirac distribution,

$$f_0\left[\epsilon(\mathbf{p})\right]=\left[1+e^{\frac{\epsilon(\mathbf{p})-\epsilon_F}{k_B T}}\right]^{-1} \tag{1}$$

Where the Dirac electrons obeying the linear dispersion relation $\epsilon(\mathbf{p})=\hbar v_F\,|\mathbf{p}|$ ($\hbar$ is the Planck constant), with $\mathbf{p}$ and $v_F$ denoting momentum and Fermi velocity, respectively, $\epsilon_F$ for Fermi energy, $k_B$ the Boltzmann constant, and $T$ for temperature. In presence of the THz pulse, the driven transient state is characterized by the distribution function $f(t,\mathbf{p})$, the time-dependent evolution of which is governed by the time evolution equation [39,40]:

$$\frac{\partial f(t,\mathbf{p})}{\partial t}=\frac{e}{\hbar}E(t)\nabla_{\mathbf{p}}f(t,\mathbf{p})+\frac{f_0(\mathbf{p})-f(t,\mathbf{p})}{\tau}, \tag{2}$$

where $e$ is the electron charge, and $\tau$ is the scattering time. The incident field $E(t)$ is defined with the experimental THz field.

To get a better insight into the THz nonlinearity of 3D Dirac semimetal, THz-driven nonlinear kinetics and time-resolved distribution function were calculated and plotted in Figure 3. The electric field of the linearly polarized pump pulse is set along the $\mathbf{p}_z$ direction. The high-quality nanotip-$Cd_3As_2$ with a gap of 50 nm under the all-optimized excitation and collection conditions, is consistent with experimental conditions. The transient-state distribution functions corresponding to the four typical time delays are illustrated in Figure 3a-d. The 3D illustration in Figure 3a-d intuitively describe the process of electron density being stretched over a timescale of approximately ~20 ps and the process of electrons oscillating near the Dirac cone ($\mathbf{p}_z=0$) in momentum space. This 3D illustration is closely correlated with the distribution function $f(t,\mathbf{p})$ (Equation (4)). For clarity, the diagrams of the near field coupling schemes and the equilibrium electron density in the upper band under the pump of 0.5 THz are sketched above the calculated curves, respectively.

As depicted in Figure 3a, the electrons in the upper band are in thermodynamic equilibrium in the initial state, populating the Dirac cone to approximately the Fermi energy level, according to the Fermi-Dirac distribution. In the presence of the THz pulse, the electrons are accelerated and oscillated by the THz electric field (Figure 3b and c). Although this transient process dissipates rapidly, the repeated acceleration and oscillation of the electron dynamics induced nonequilibrium electron distribution in the upper band.

Especially, at the peak of the THz field (Figure 3b), the electron distribution is most strongly mobilized resulting in the maximum current density (Figure 3b). Especially, at the peak of the THz field (Figure 3b), the electron distribution is most strongly mobilized resulting in the maximum current density (Figure 3b). According to equations 4-7, we can conclude that the evolution of the current density induced the efficient THz nonlinearity generation. Finally, at the end of the driving THz pulse (Figure 3d), the electron distribution returns to quasi-equilibrium state under low peak power THz field.

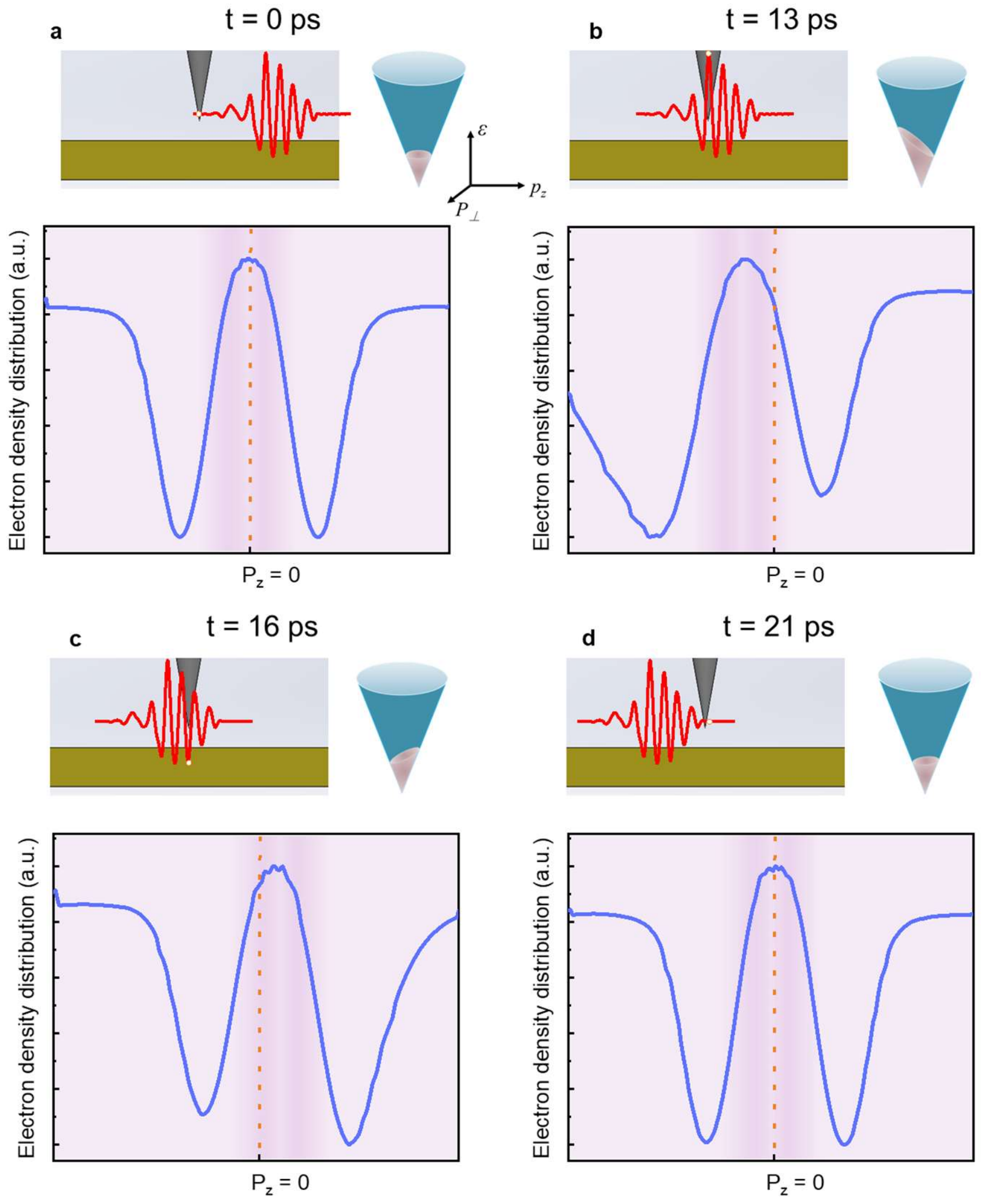

**Figure 3. Nonequilibrium intraband dynamics driven by THz pulses.** a-d Electron distribution in the upper band of the Dirac cone corresponding to various time-delays ~20ps. Upper figures in a-d: illustrations of the near field coupling of the nanotip apex – $Cd_3As_2$ system when the pump pulse arrives. 3D illustrations of the distribution function $f(t,\mathbf{p})$.

After we obtain transient distribution $f(t,\mathbf{p})$ by solving the time evolution equation, the current density of the Dirac electron distribution can be calculated as

$$J_D(t) = ev_F \int \frac{d\mathbf{p}}{(2\pi)^3} \hat{\mathbf{p}} f(t,\mathbf{p}) = J_\omega(t) + J_{3\omega}(t) + \cdots\cdots \tag{3}$$

where $\hat{\mathbf{p}}$ denotes the unit vector along the momentum direction. Note that the equation is symmetric under rotations in the $k_y k_z$ plane, only the electron density in $k_x$ was calculated. Here we focused on the induced nonlinear current of $3\omega$,

$$J_{3\omega}(t) = \sigma(3\omega)\left[E(t)\right]^3 e^{-3i\omega t} \tag{4}$$

where $\sigma(3\omega)$ is the conductivity at the third harmonic. Finally, the third harmonic electric field can be given (further details can be found in extended Figure S6):

$$E_{3\omega} \propto \frac{dJ_{3\omega}(t)}{dt} \tag{5}$$

$Z_0$ is the impedance of free space.

To further characterize intensity evolution of the third harmonic generation, we calculated the near-field $Cd_3As_2$ response under different localized field enhancement effects. In the experiment, we modify the localized field enhancement effects by adjusting the nanotip-sample distance (Figure 4 a-c). Figure 4d illustrates the induced dipole moment $\boldsymbol{p}$ and the localized field enhancement evolution trends by adjusting the nanotip-sample distance. The quantitative relationship between dipole moment and the distance can be determined by:

$$E(t) \propto \boldsymbol{p}(t) = \left| \iint_S \sigma_S(t) z dS \right| \tag{6}$$

$$\frac{d}{dt}\boldsymbol{p}(t) = \int d^3x' \boldsymbol{J}\left(x', t\right) \tag{7}$$

Where $E(t)$ is the amplitude of the tip-localized field, $\sigma_S$ is the surface charge density, $dS$ is the surface element and $z$ its vertical position, $\boldsymbol{J}$ is the current density, $x'$ is the horizontal positions.

When the nanotip-sample distance is 50 nm, the localized electric field strength $E(t)$ at the tip apex is ~350 kV/cm, reaching about six times the incident electric field strength (~60 kV/cm) (see extended Figure S7). Adjusting the local electric field intensity at the tip apex interacted with $Cd_3As_2$ film through the nanotip-sample distance, the experiments and simulations to investigate the dependence of the THz nonlinearity on the localized field intensity were conducted.

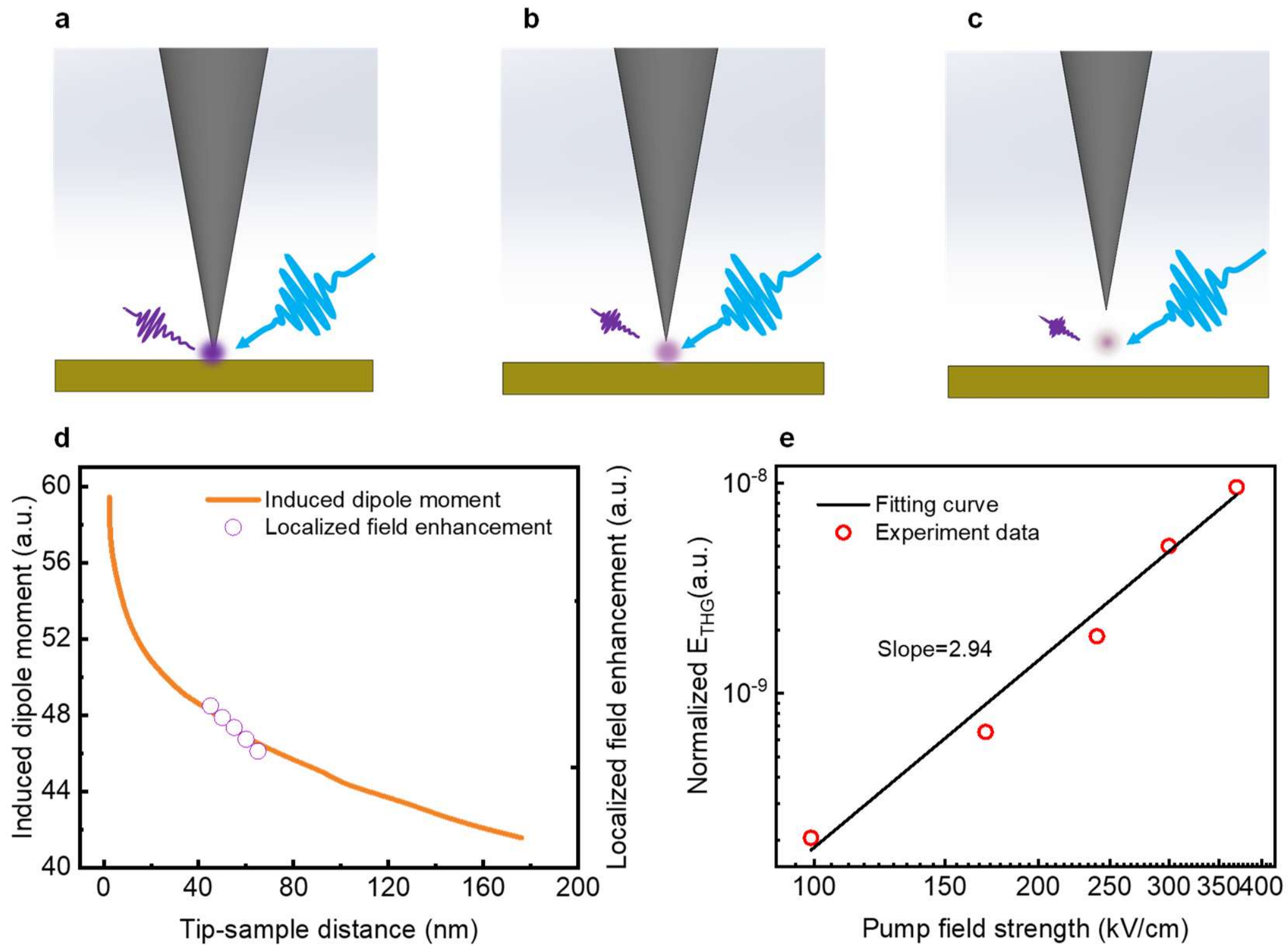

**Figure 4 Dipole moment, localized field enhancement and the harmonic intensity induced by localized THz pulses in near field.** a-c Illustrations of the near field coupling intensity evolution corresponding to the tip-sample distance. d Dependence of the near-field induced dipole moment and the localized field enhancement on tip-sample distance. e Dependence of normalized third harmonic generation $E_{THG}$ on localized electric field.

As summarized in Figure 4e, the fluence dependence of the third harmonic radiation remarkably obey the cubic law but exhibits a power-law dependence as $I_{3f} \propto I_f^{2.94}$ on the pump-pulse intensity which reveals an expected nonlinear response of THG. Clearly, the efficiency and fluence dependence of the THz nonlinearity are determined by the localized field intensity. Refined near-field modulation enhances the local field strength at the nanotip apex and the intensified local driving field further stretches the transient distribution of Dirac electrons, thereby enhancing the generation of THz nonlinearity. That is, the microscopic origin of THz nonlinearity can be attributed to the nonlinear dynamics of the electron distribution excited by the localized THz field in nanotip-$Cd_3As_2$ gap.

**Near field third harmonic imaging with nanoscale resolution**

The method combining the high peak power THz source [41,42] and the THz s-SNOM system, allows the synchronous measurement of near-field linear and nonlinear responses of novel materials with a nanoscale resolution [22,43,44]. Figure 5a illustrates an AFM morphology (5 μm × 5 μm) of the $Cd_3As_2$ surface (black part) with triangular sapphire particles (white part). Figures 5 b and c are THz near-field scanning imaging at 0.5 THz and 1.5 THz, corresponding to conventional and nonlinear THz near field mappings, respectively. Clearly, the image contrast of AFM topography and THz s-SNOM is opposite for the same $Cd_3As_2$-sapphire interface, originating from the distinct sampling methods of these two techniques. The AFM directly records the feedback voltage signal for the fixed nanotip height at the z-direction to reflect the surface morphology. For the sample consisting of the bottom $Cd_3As_2$ film and the top sapphire particles, the feedback voltage signal of the $Cd_3As_2$ surface is lower than the sapphire particle. In contrast, the THz s-SNOM records near-field scattering signals between the nanotip and the sample can be modulated the surface morphology and near field coupling. Due to the dielectric constant of $Cd_3As_2$ is much higher than sapphire, the coupling between tip-$Cd_3As_2$ is much strengthener than that of tip-sapphire. Therefore, the near-field scanning locations with $Cd_3As_2$ has a more remarkable scattering THz signal.

Then we demonstrate the advantage of nonlinear imaging in improving the spatial resolution for THz s-SNOM. Figures 5 d and e depict the 1st derivative curves of line cut profile of the transition region between $Cd_3As_2$-sapphire labeled in Figure 5a. The nonlinear harmonic imaging possesses higher resolution comparing with linear scheme. Moreover, both the $Cd_3As_2$ and Sapphire can response to the fundamental scattering THz waves, while only the $Cd_3As_2$ can be excited and emit third harmonic. Therefore, it can be obtained the a near-field scattering amplitude ratio ($R = S_{Cd3As2}/S_{sapphire}$) of the linear near-field imaging ($R = 3.6$) is much lower than that of nonlinear near-field imaging ($R = 45$). Here the spatial resolution of near field nonlinear THz imaging is about 200 nm ($\lambda/3000$), illustrating nanoscopic near field THz THG of 3D Dirac semimetal are achieved (The simulation result is shown in extended Figure S8). The almost same resolution of 0.5 THz and 1.5 THz can be attributed to the resolution of scanning probe microscopy is mainly determined by the dimensions, height and control accuracy of the tip. Note that here we just combine low-cost educational AFM (EDU-AFM1/M, Thorlabs) and home-made piezoelectric control system to obtained tapping mode for temporal synchronization between tip oscillation and incident THz wave, constraining the spatial resolution of THz s-SNOM. Providing the advanced high-precision AFM with tapping mode were adopted, the spatial resolution of nanometers and the field strength of MV/cm (see extended Figure S7) can be expected.

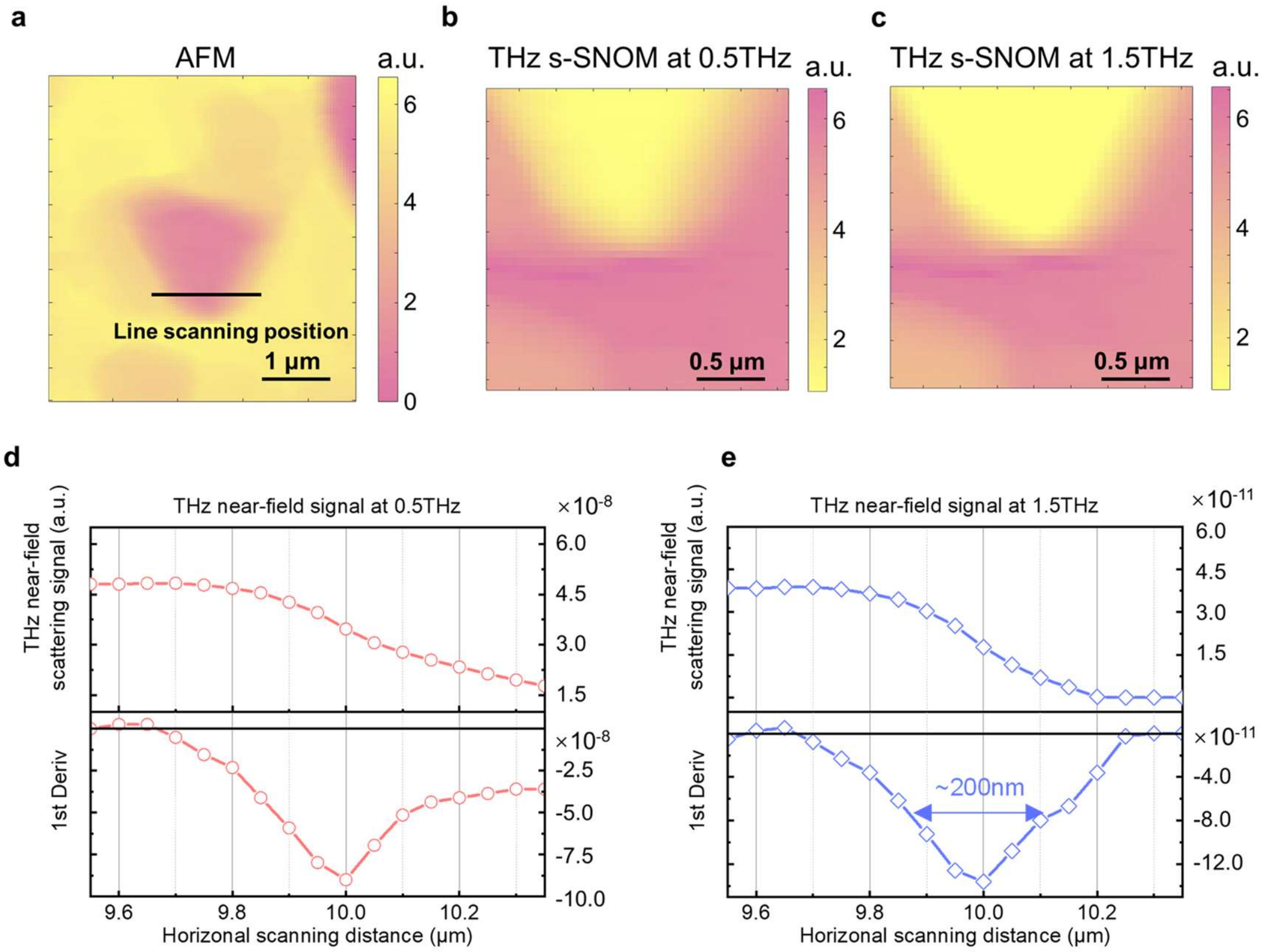

**Figure 5. Near field third harmonic imaging with nanoscale resolution.** a AFM imaging results. The central pink triangular part is a sapphire particle, and the surrounding yellow areas are $Cd_3As_2$ sample regions. b THz s-SNOM image. The intensity data in the image is extracted from the near-field scattering 0.5 THz power spectra. c THz s-SNOM image. The intensity data in the image is extracted from the near-field scattering 1.5 THz power spectra. d 0.5THz near-field optical signal profiles along the black lines in a. e 1.5THz near-field optical signal profiles along the black lines in a.

## Potential applications of intense THz s-SNOM

Although the THz s-SNOM has rapidly developed and unlocked the ability to resolve nanometers in spatial, the THz nonlinearities excited by high-power THz pulses still remain in the far field with spatial resolution of millimeter. Here the intense THz s-SNOM, combing high-power THz pulses with s-SNOM system, achieves dual advantages of high spatial resolution (~100 nm) and high-intensity excitation field (~300 kV/cm) (Figure 6), thereby providing a promising technique to explore the THz nonlinearity, carrier dynamics and quantum coherent physics (Superconductor modes, Rydberg states, etc.) in condensed matter.

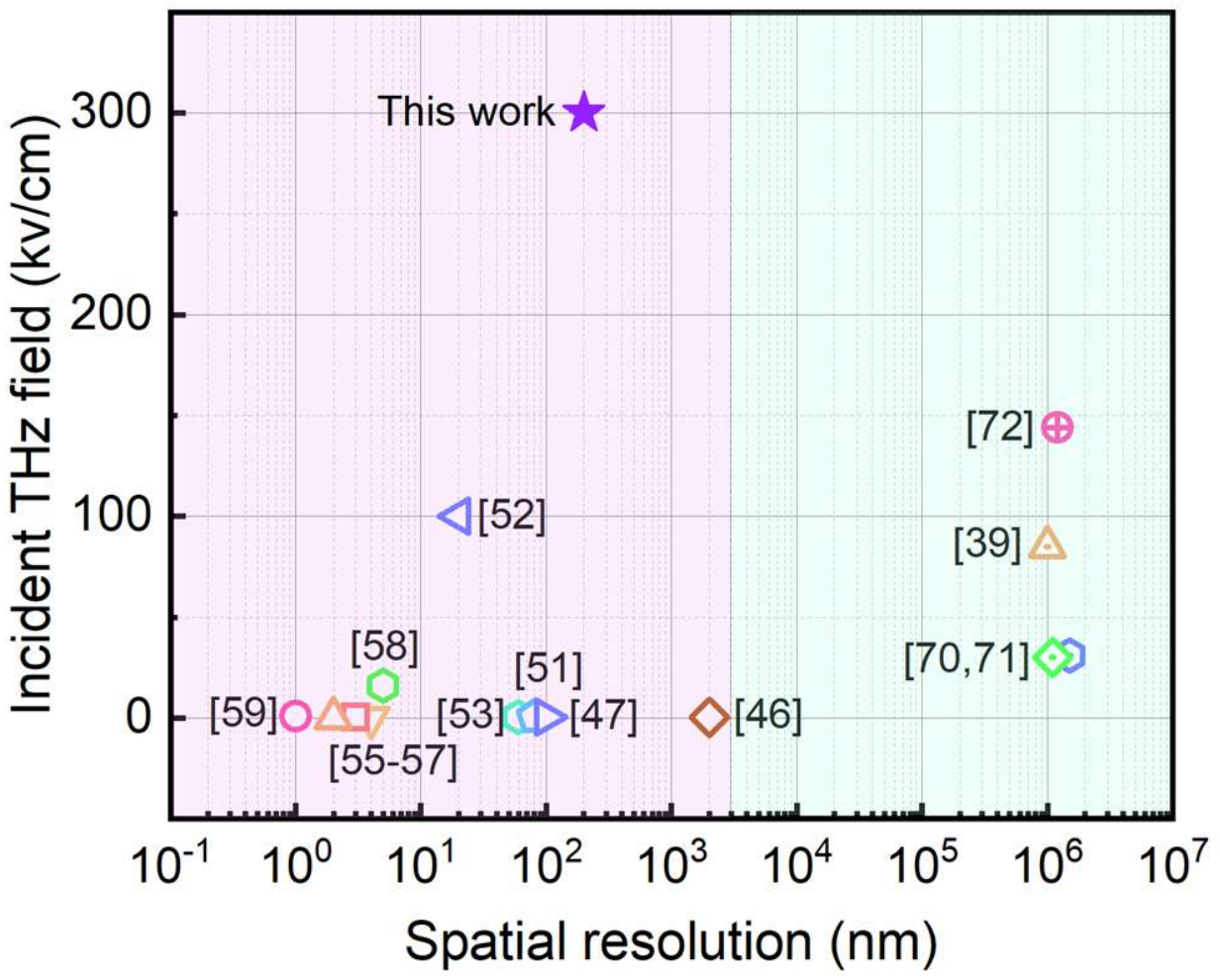

**Figure 6 Spatial resolution and field strength of the THz near field system (purple area) and the THz nonlinearities excited in far field (blue area).** THz s-SNOM[46,47,51,53]; IR nanoscopy[52]; THz STM[56–59]; THz nonlinearities generation in far field[39,71-72].

Firstly, for the currently discovered and confirmed superconductors, including cuprates, iron pnictides, $NbSe_2$, and $MgB_2$, the gap values and the relevant collective modes are in the THz range [60,61]. It has been confirmed that THz s-SNOM has the potential for observing and modulating the superconductor modes in nanoscale such as superfluid plasmons, amplitude modes, Bardasis-Schrieffer modes, and Carlson-Goldman modes. Therefore, the intense THz s-SNOM can be used for detecting collective modes and provide more insights into both the excitation and relaxation dynamics of superconductors [62,63]. Furthermore, the tunable pump wavelength can adapt to the deep sub-band-gap and coherently excite THz phonon [64]. The nanomanipulation induce the extremely small spatial lattice displacement driven by transient intense THz field can induce broken symmetry coherent phonons and brighten the dark excitons states. Therefore, the intense THz s-SNOM can be used to explore symmetry-selective coupling of THz-induced phonon mode and the fine band structure of exciton or Rydberg states [65] [66]. Additionally, the intense THz s-SNOM can also performed at high-intensity magnetic field and ultralow temperatures to map the breaking of time-reversal symmetry and magneto-optical properties at the nanoscale [67,68]. As a frontier technique directly revealing the dynamic evolution of carrier and quasi-particles in quantum system [19,69], the nanomanipulation can catch the transient interference fringes, near-field patterns and the dispersion of polaritons[18], and manipulate the hyperbolic Rydberg polaritons.

## CONCLUSION

In summary, we constructed an intense THz s-SNOM for coupling the high peak power THz field into near-field for producing near-field THz nonlinearity with nanoscopic resolution. The nanoscale strong near-field coupling and manipulation accurately revealed the nonlinear response of Dirac semimetal $Cd_3As_2$ to the strong THz field. The effective nonlinear conversion efficiency of $Cd_3As_2$ film due to the localized field enhancement near nanotip has been revealed by near-field scattering THz spectra. The transient Dirac electron dynamics in the upper band of the $Cd_3As_2$ have been further calculated to describe the Dirac electron distribution evolution and analyze the physical scenario of THz nonlinearity. Meanwhile, the nonlinearity of THG has been demonstrated by the power-law dependence of the THz harmonics. Finally, the near-field third harmonic imaging with resolution of 200 nm ($\lambda$/3000), that is nanoscopic near field THz THG of 3D Dirac semimetal have been achieved. The high peak power THz s-SNOM can provide a great platform for exploring and manipulating the nonlinear physics, carrier dynamics and quantum coherent phenomena in condensed matter driven by the localized THz field with nanoscale resolution, thereby guiding the development of the integrated high-performance nonlinear photonic devices.

**Acknowledgements** This work is supported by the funding support from National Natural Science Foundation of China (W2412044), Russian Science Foundation (25-49-00154) and Fundamental Research Funds for the Central Universities (63253214).

**Author contributions** Weiwei Liu lead the research. Zeliang Zhang and Xingyou Li performed the experiment. Zeliang Zhang and Pengfei Qi performed theoretical calculations. Weiwei Liu, Zeliang Zhang, Pengfei Qi, See Leang Chin, and Pierre Agostini contributed to interpreting the results. Weiwei Liu conceived the original idea, conceptualized the study, designed and supervised the overall research project, secured the necessary financing, and was responsible for project administration. All authors contributed to discussing the results and manuscript revision.

**Competing interests** The authors declare no competing interests.

**Correspondence and requests for materials** should be addressed to Weiwei Liu.

## Methods

**Basic experimental set-up of the THz pulse generation and detection**

The THz pulse is produced by the two-color laser filamentation. Laser pulses (45 fs, 4.4 mJ, 800 nm) with horizontal linear polarization were focused by a plano-convex lens of 50 cm focal length. A 107 μm thick β-BBO crystal with type Ⅰ phase matching was used to generate the second harmonic wave (SH). We have further controlled the polarization and compensated the time delay of fundamental wave (800 nm) and second harmonic pulses (400 nm) through double wavelength waveplate (DWP) and α-BBO crystal to precisely modulate the THz pulse intensity. The off-axis parabolic mirror (OAP) with 2 inches diameter and the 4 inches reflected focal length was used to collimate THz pulses. The collimated THz pulse is focused on the near-field system. As photoconductive antenna is used to detected the near-field scattering signal. The lock-in amplifier (Stanford Research Systems SR830) is used in our system to improve SNR.

**Basic experimental set-up of the THz s-SNOM system**

The near-field system is an improvement over an atomic force microscopy (AFM) system, and three key ideas guide the signal extraction process: 1) maintaining the probe's vibration frequency (near-field signal repetition frequency) different from the incident signal and background scattering signal, 2) keeping the probe's vibration synchronized with the pulsed THz source during the vibration process, and 3) using a lock-in amplifier to extract near-field scattering signals (extended Figure S2).

The basic component controlling the probe's vibration is a PZT (2×2×2mm). Independently controlling PZT vibration with an external power supply or signal generator to synchronize it with the THz pulse is challenging. Hence, a synchronization signal generator (SDG) is employed to modulate the trigger signal of the laser pockel cell. It converts a 500Hz electrical pulse signal into a synchronized 20Hz electrical pulse signal (extended Figure S3 shows the connection of the synchronization signal). The timing of the pulse reaching the sample position is detected by placing a photodetector (PD) at the sample position. A digital delay signal generator (DG535) is used to control the timing of the 20Hz electrical pulse signal and the PD detection signal until the 20Hz electrical pulse can completely cover the PD detection signal. This 20Hz electrical pulse signal directly controls the piezoelectric ceramic's vibration, ensuring that the probe's vibration timing remains consistent with the THz pulse timing. When the probe vibrates close to the sample, fulfilling the near-field conditions, the THz pulse can accurately reach the near-field region. The output modulation signal of DG535 is synchronized with the lock-in amplifier as the reference frequency, enabling second-order demodulation to obtain the near-field scattering signal.

**Full Wave Numerical Simulation**

The full wave numerical simulation is used to calculating the near-field scattering efficiency and the localized field enhancement. The simulation of the scattering nanotip is based on the realistic experiment parameters, such as the geometrical parameter of the nanotips, the distance between the nanotip apex and the sample surface and the incident THz field. The nanotip is modeled as conical frustum and a semispherical apex with geometric parameters shown in extended Figure S4. The influence of the cantilever is neglected. The Dirac semimetal $Cd_3As_2$ film (thickness: 100nm) is defined by the Drude model. The complex dielectric constants and the complex conductivity of the Dirac semimetal $Cd_3As_2$ is proposed in Reference [29]. The dielectric constant of the sapphire in this article is 8.7 (Reference [29]). The incident field $E_{inc}$ is defined as a p-polarized plane wave, incident at 60° relative to the surface.

The incident field induce a net vertical dipole $\boldsymbol{p}$ which is interacted by the nanotip and the Dirac semimetal $Cd_3As_2$ film. The near-field scattering amplitude $E_{sca}$ can be considered as the radiation of this net dipole and is proportional to the amplitude of the dipole moment $\boldsymbol{p}$. The near-field dipole moment is proportional to the surface charge density $\sigma_S$. The near-field scattering field can be described as:

$$E_{scat} \propto \boldsymbol{p} = \left| \iint_S \sigma_S z dS \right|$$

where $dS$ is the surface element and $z$ is the vertical position. The equation describes the integral of the surface charge density along the $z$ direction form each cross section of the nanotip.

The electric dipole moment of a charge distribution is defined by the integral expression:

$$\boldsymbol{p}(t) = \int d^3 x' \sigma_S \left( x', t \right)$$

In equation (1), $x'$ is the horizontal positions. $\sigma_S \left( x', t \right)$ determines the charge density of the nanotip' s apex surface. Charge conservation relates the charge density to the current density by the continuity equation [21,22]:

$$\nabla \cdot \boldsymbol{J} + \frac{\partial \sigma_S}{\partial t} = 0$$

$\boldsymbol{J}$ is the current density. The following relation between the electric dipole moment and the current density follows ($\mathbf{D}$ is the electric displacement vector, $H$ is the magnetic field intensity, and $\omega$ is the angular frequency):

$$\frac{d}{dt} \boldsymbol{p}(t) = \int d^3 x' \boldsymbol{J} \left( x', t \right)$$

$$\boldsymbol{J} = \begin{vmatrix} \mathbf{i} & \mathbf{j} & \mathbf{k} \\ \frac{\partial}{\partial x} & \frac{\partial}{\partial y} & \frac{\partial}{\partial z} \\ H_x & H_y & H_z \end{vmatrix} - (-i\omega \mathbf{D})$$

As it is shown in equation (4), the dipole moment is proportional to the current density. Therefore, we can estimate the near-field scattering intensity by approximating the surface current density.

**Estimation of THz field strength by photoconductive antenna and Golay cell**

Estimating the THz field strength produced by the two color laser filament is supposed to measure the pulse energy, pulse duration (time domain curves), and focusing spot size. The pulse energy is measured by the Golay Cell, and the optical responsivity is shown as:

$$P_{THz} = \frac{V_{THz} \times \frac{P}{U}}{T_{filters}}$$

$$E_{THz} = \sqrt{2\eta \cdot 2 \cdot \frac{\varepsilon_{THz}}{\tau \cdot \pi \cdot \left(\frac{d}{2}\right)^2}}$$

$P_{THz}$ is the average THz power, $V_{THz}$ is the measurement result of the Golay cell, $P/U = 2\times10^{-5}$ W/V is the optical responsivity parameter of the Golay cell, $T_{filters}$ is the transmissivity of THz waves transmitted through air and filter, $E_{THz}$ is the THz field strength, $\eta = 377$ is impedance, $\varepsilon_{THz}$ is the single-pulse energy of THz pulse, $\tau$ is the pulse duration of the THz pulse, $d$ is the focusing spot diameter of the THz pulse.

The time domain curve of the THz pulse is measured by the photoconductive antenna, and the experimental result is shown in the extended figure 1, which shows the pulse duration of the THz pulse is 0.5 ps. The THz beam diameter can be measured by eclipsing the THz spot with a semi-infinite mask while monitoring signal variation (knife-edge technique). In the experiment, the THz beam is focused by an off axis parabolic mirror, and the THz beam spot diameter is 1.5mm.

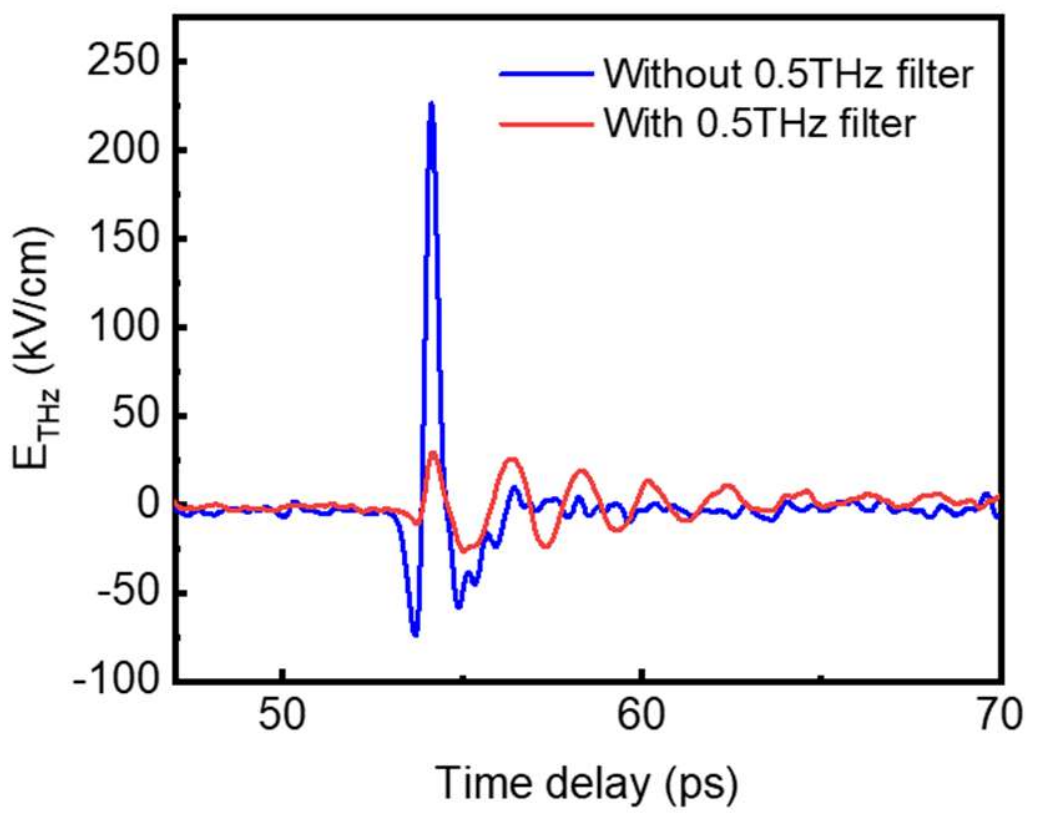

**Extended Figure S1. THz electric field time domain curve.** The blue line is the THz radiation generated by the two-color laser filament without 0.5THz filter. The red line is the THz radiation with 0.5THz filter.

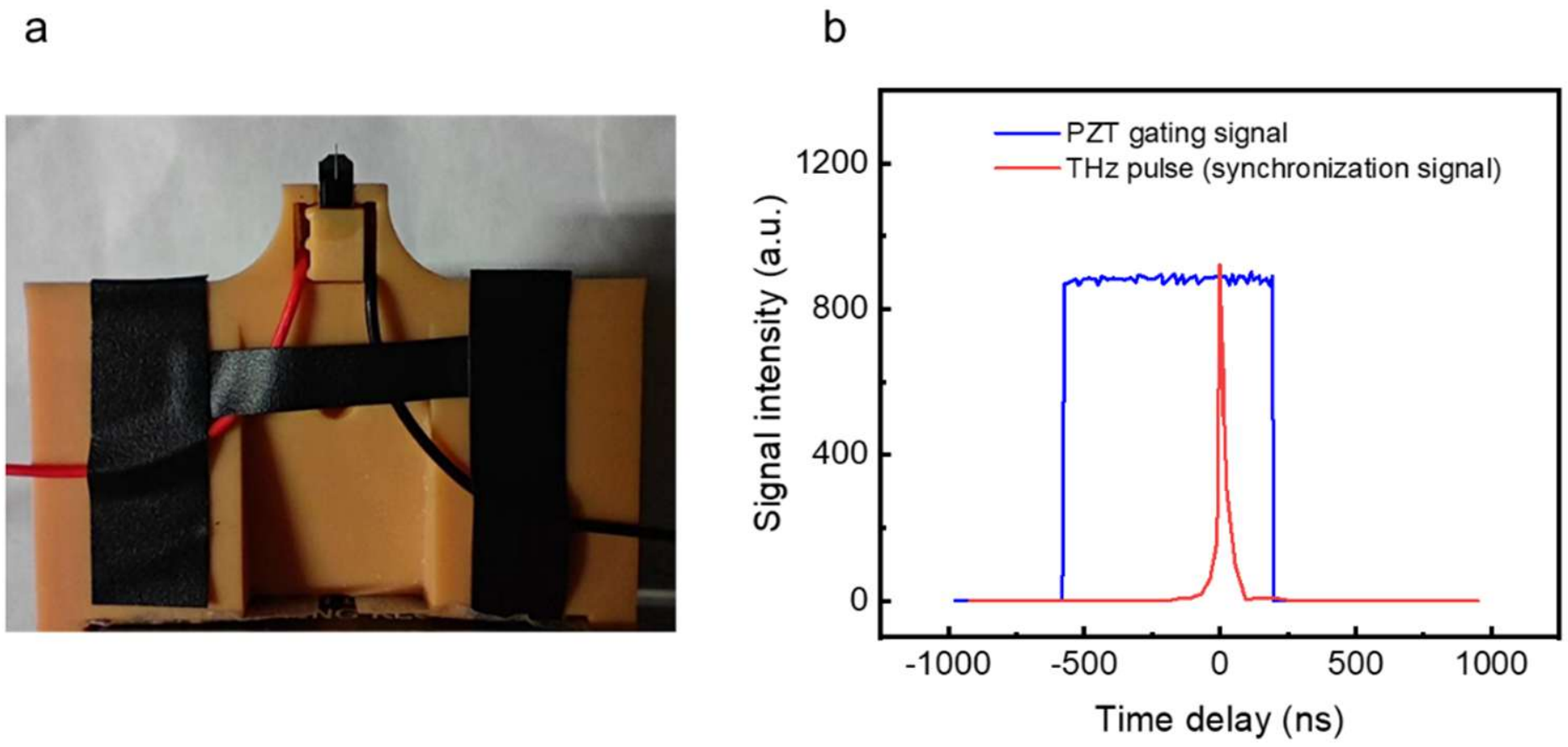

**Extended Figure S2. Experiment setup.** a piezoelectric ceramics feedback setup for modulating the vibration of the nanotip. b the blue line shows the vibration signal of the nanotip and the yellow line shows the synchronous timing signal of the THz pulse.

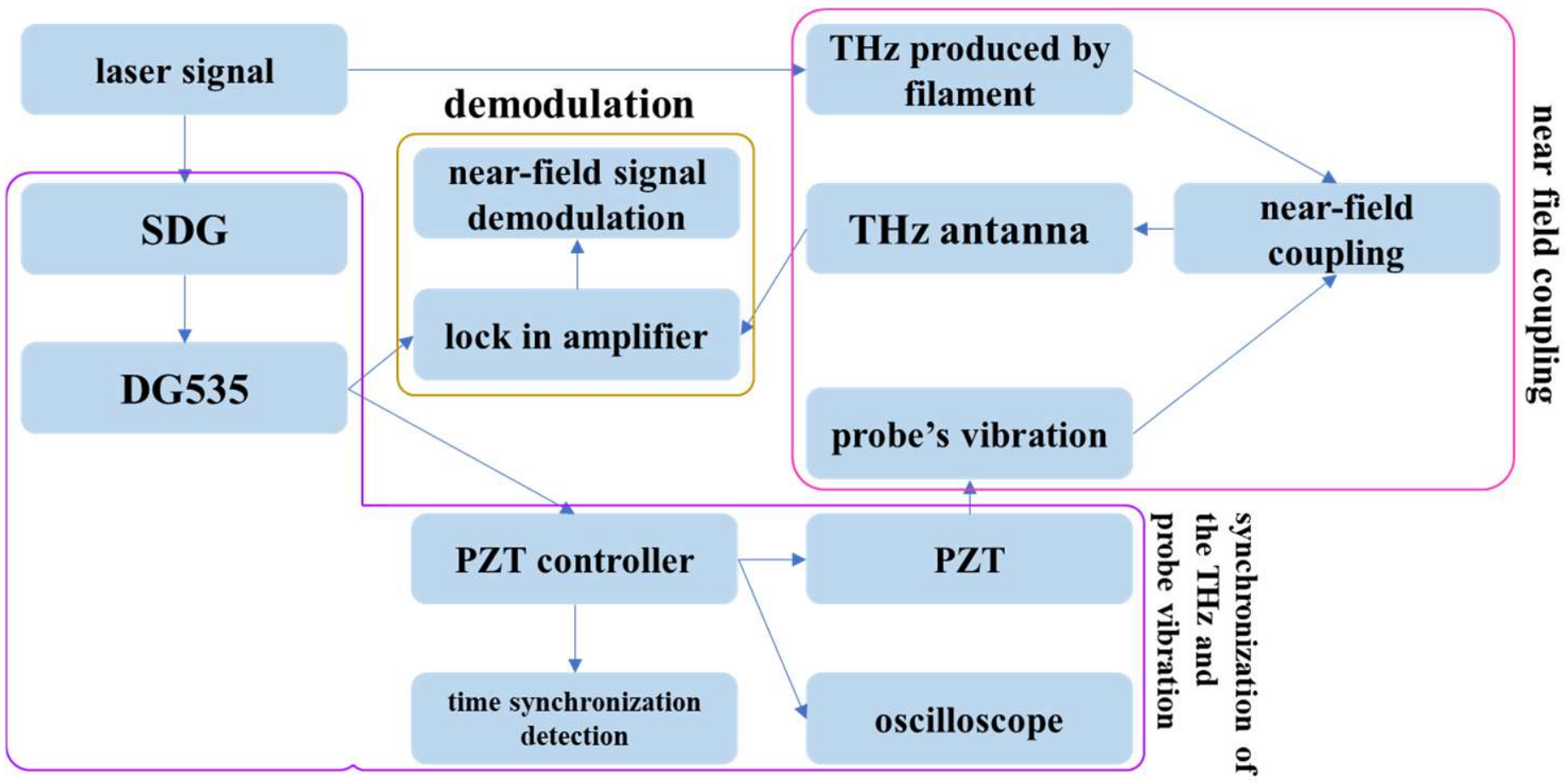

**Extended Figure S3. Circuit system connection.**

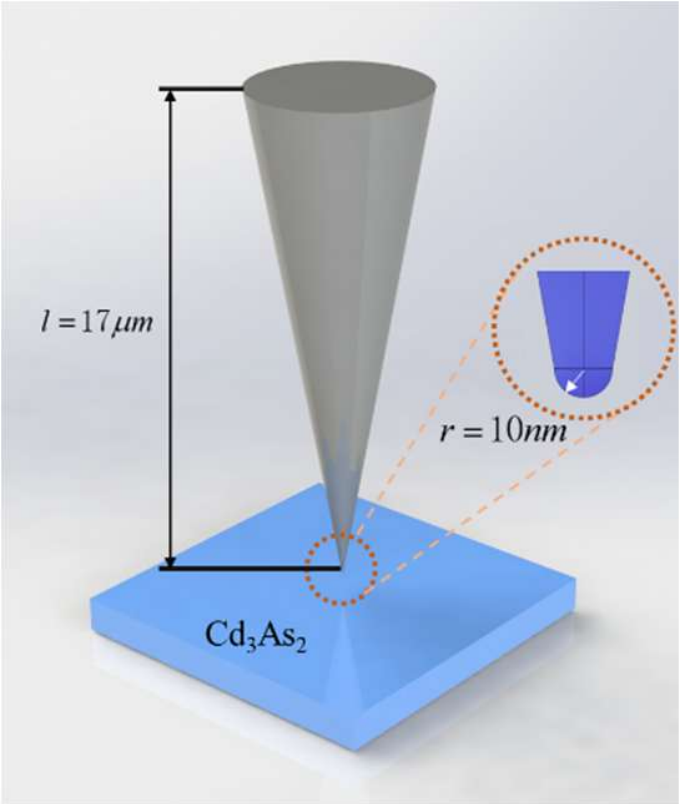

**Extended Figure S4. Sketch of the numerical model**

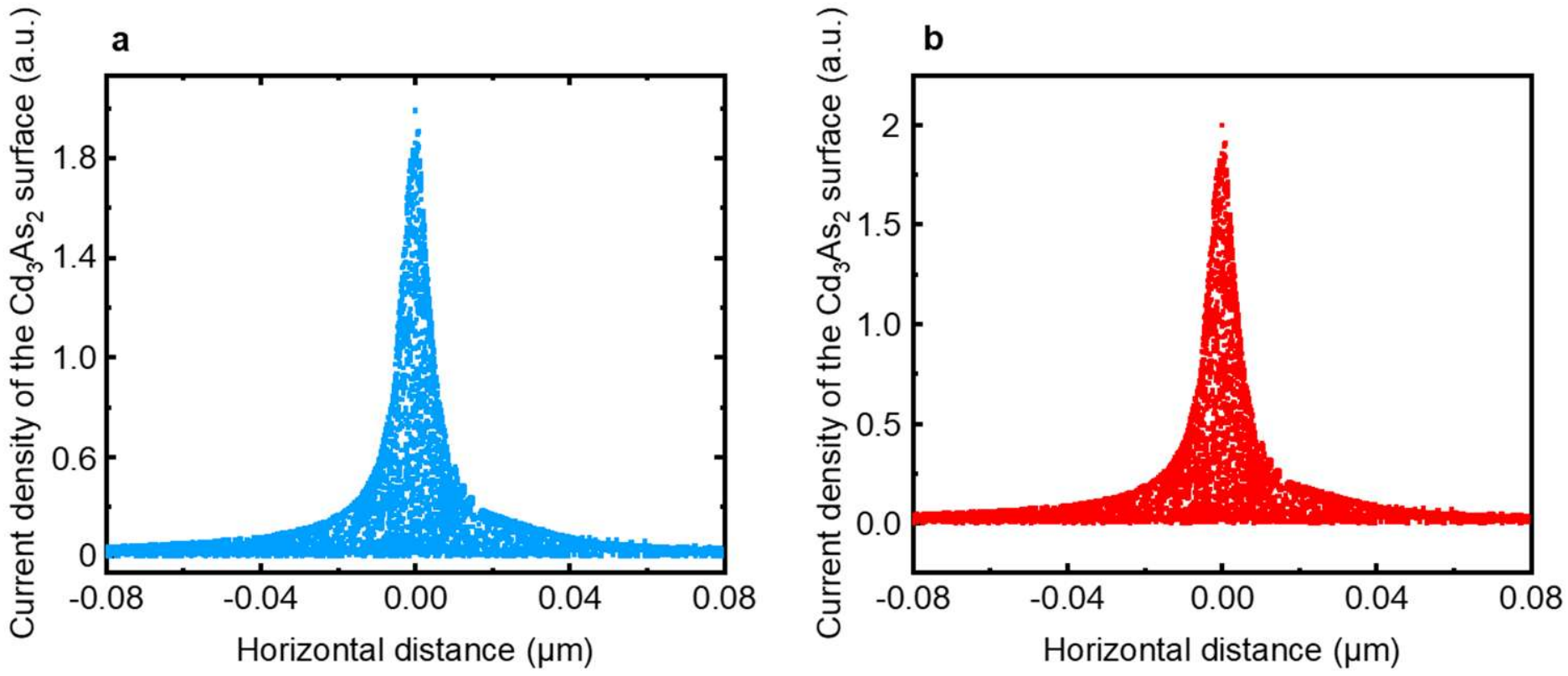

**Extended Figure S5. Surface current density of the $Cd_3As_2$ surface under the nanotip apex.** The distance between the nanotip apex and the $Cd_3As_2$ surface is 50nm. a The current density distribution of the $Cd_3As_2$ surface with 1.5THz. b The current density distribution of the $Cd_3As_2$ surface with 0.5THz.

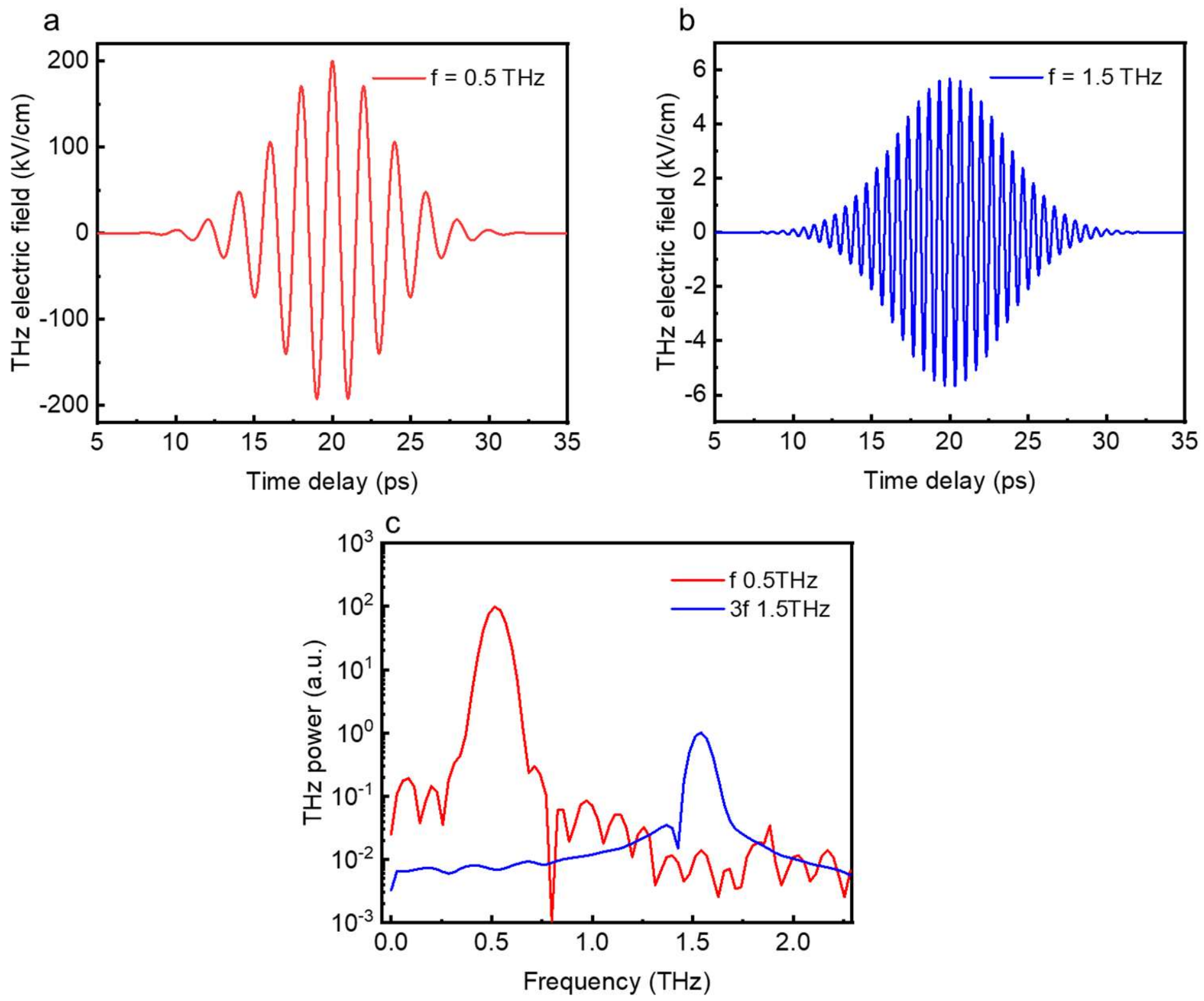

**Extended Figure S6. THz electric field calculation of the dynamic evolution of the Dirac electrons model.** a Fundamental pump wave b Third harmonic wave. c the spectra of a and b.

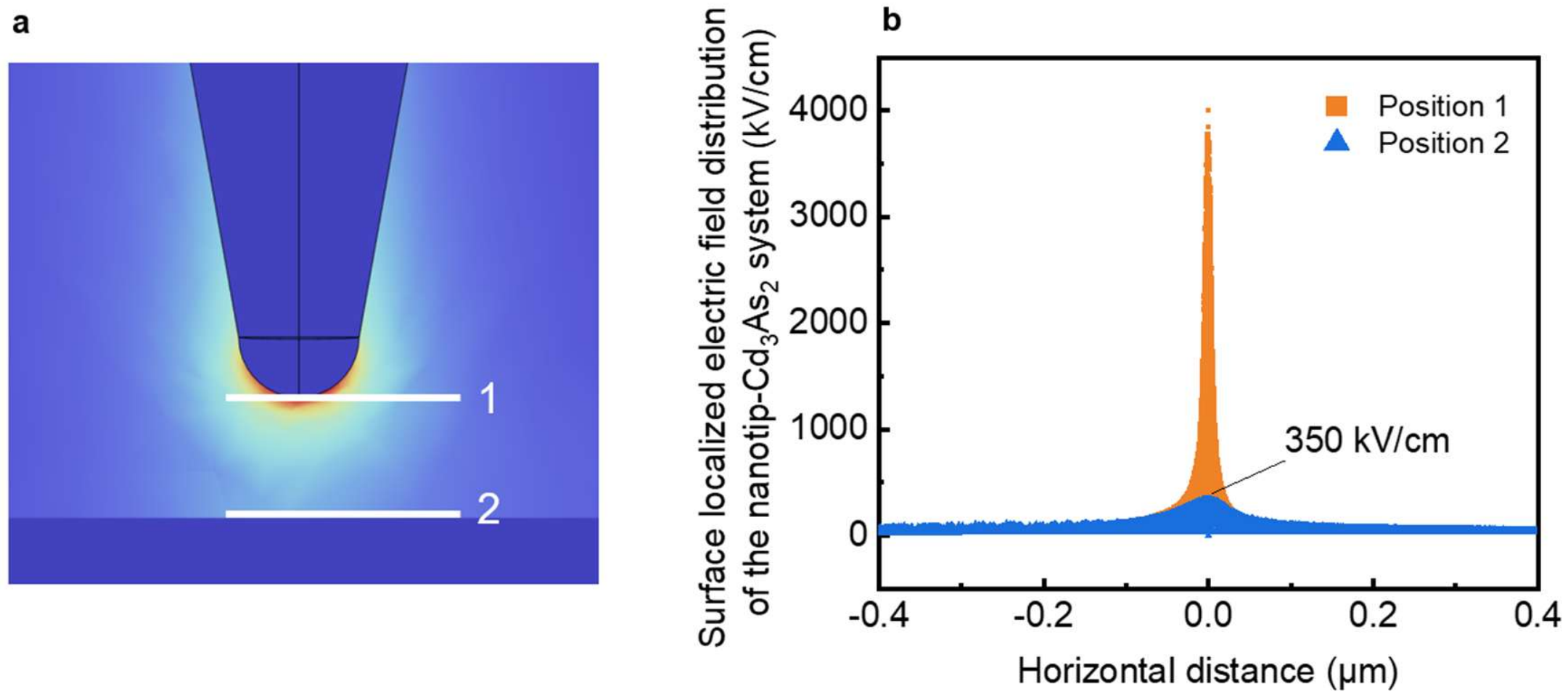

**Extended Figure S7. Localized electric field intensity distribution. Position 1 shows the distance of between the test surface and the nanotip apex is 0nm. Position 2 shows the distance of between the test surface and the nanotip apex is 50nm**. a Schematic diagram of the local

electric field distribution of the nanotip apex – $Cd_3As_2$ system. b Localized electric field intensity at position 1 and position 2, respectively.

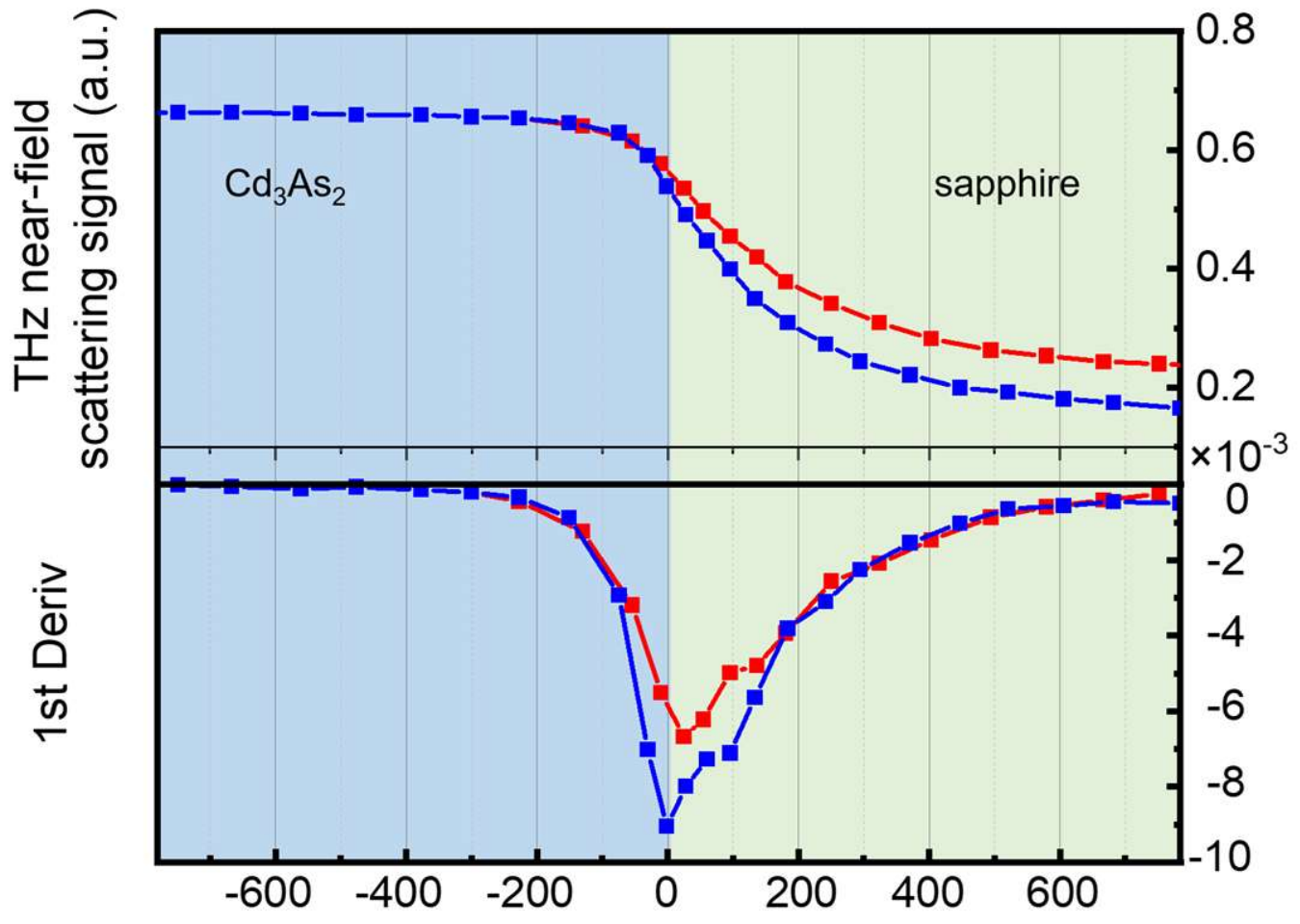

**Extended Figure S8. Near field scattering signal of horizontal scanning across the $Cd_3As_2$ - sapphire.** The red line is the near-field scattering signal of 0.5THz and the blue line is the near-field scattering signal of 1.5THz.